\documentstyle[11pt,paspconf]{article}

\begin{document}

\title{Type Ia Supernovae and ${\bf q_0}$\footnotemark[1]}
\footnotetext[1]{To appear in
`Supernovae and Cosmology' eds. L. Labhardt, B. Binggeli, R. Buser,
Basel: University of Basel}
\author{Bruno Leibundgut}
\affil{European Southern Observatory, Karl-Schwarzschild-Strasse 2, \\D-85748
Garching, Germany}

\begin{abstract}
Type Ia Supernovae hold great promise to measure the cosmic deceleration.
The diversity observed among these explosions, however, complicates their
ability to measure cosmological parameters considerably. The comparison of
near and distant objects depends critically on the 
global properties of the samples. Systematic errors are introduced in
cases where the samples are not compatible.
Our discussion of systematic uncertainties
emphasizes the dependence on the zero-point defined by the nearby
SNe~Ia sample and the influence of the application of light curve
corrections to the luminosity. Other error sources are in principle
controllable. We illustrate the effects in an analysis of four SNe~Ia at
$z > 0.4$.

\end{abstract}

%\keywords{Cosmology, Supernovae, Distance Indicators}

\section{Introduction}

It is more than 70 years since it was realized that the Universe is expanding,
yet we still do not know its exact dynamic state.
The long debate on the present-day value of the
expansion rate, Hubble's constant, appears to be coming to an end and most
astronomers agree on similar values. The coming years will see a
discussion on the global change of the expansion rate and the final
fate of the Universe. Supernovae have been at the forefront of this
research for the last decade and will also play an important role in
the near future by measuring the deceleration $q_0$ and separating
the contributions of matter $\Omega_M$ and a cosmological constant 
$\Omega_{\Lambda}$.

'Supernovae Type~Ia [are] the best standard candles known so far' (Branch \&
Tammann 1992).
Even though this statement has been challenged in recent years
it holds pretty much true. The magnitude-redshift relation of Type Ia
Supernovae (SNe~Ia)
has always displayed a rather small scatter (Kowal 1968, Tammann \&
Leibundgut 1990) and almost all SNe~Ia exhibit very similar $B$, $V$,
and $I$ light curves (Leibundgut et al. 1991, Hamuy et al. 1996). 
A debate on SN~Ia uniformity has started with
the discovery of a few under-- and overluminous events 
(SN~1991bg: Filippenko et al. 1992a,
Leibundgut et al. 1993, Turatto et al. 1996; SN 1992K: Hamuy et al. 1994;
SN~1991T: Filippenko et al. 1992b, Phillips et al. 1992; SN~1992bc and 
SN~1992bo: Maza et al. 1994)
which show many, but not all,
signatures of what normally is called a SN~Ia.
There are important differences in the appearance of these objects,
but it seems that a clear separation into
subclasses is not possible, if they are indeed distinct
from the majority of
SNe Ia. Even highly selected samples (Vaughan et al.
1995, Sandage et al. 1996, Saha et al. 1997) show
internal scatter in absolute peak luminosity.

It is the maximum luminosity of SNe~Ia
which is normally used when they are employed as standard candles. 
A limited and small range of peak brightness is crucial.
The total scatter of nearby SNe~Ia in the Hubble flow ranges
from about 0.25 magnitudes in $B$ (Hamuy et al. 1996, Saha et al. 1997)
to 0.19 magnitudes in $I$ (Hamuy et al. 1996). 
The scatter can further be reduced by empirical corrections derived
from distance-independent parameters which
correlate with the peak luminosity.
The light curve shape is the most frequently used method
(Phillips 1993, Hamuy et al. 1996, Riess et al. 1996). 
Another option is to apply 
appropriate selection criteria to limit the range of absolute magnitudes of
known SNe~Ia in which case the pure standard candle approach may still
be valid (Vaughan et al. 1995, Sandage et al. 1996, Saha et al. 1997).

Defining subsets of objects
is a valid procedure as long as the selection does not result in a
degeneracy. None of the light curve correlations found for SNe~Ia have so 
far been explained by theory.
The use of SNe~Ia in cosmology is one of the main incentives to 
better understand SN~Ia physics.

The current knowledge of the explosion physics and the radiation
transport in the highly non-thermal envelopes is far from being
complete (Eastman \& Pinto 1993, Pinto \& Eastman 1997, Eastman 1997,
Pinto 1997, H\" oflich et al. 1993, 1995, 1996,
1997). Detailed observations obtained for many nearby objects
indicate that some of the fundamental assumptions adopted
a few years back have been in error. Especially the
assumption of a thermal radiation field in SNe~Ia is wrong and a more
exact treatment is necessary. The
emerging radiation resembles a black body in the optical fairly
well results, however, from redistribution of the flux due to the many
lines of iron and iron-like elements blocking the UV radiation. A very clear
sign of the non-thermal nature of the radiation emitted by SNe~Ia is
the time of maximum light in different filter passbands. For a simple,
expanding ball of gas in thermal equilibrium the maximum at longer wavelengths
should be reached at later phases as the gas cools
adiabatically. This is not true for most SNe~Ia where the $I$ light curve
peaks before the $B$ maximum is reached (Vacca \& Leibundgut 1996, 
Contardo et al.
1998), a sign of the wavelength dependence of the
ejecta opacity. A detailed radiation model explaining this observation
is still lacking. The spectral individuality of SNe~Ia at maximum light 
has been recognized as due to small
abundance differences in the outer envelope which are of little
relevance for the explosion physics itself (e.g. Jeffery et al. 1992,
Mazzali et al. 1993). It may be sensitive to
some of the precursor history, but not necessarily represent
a true signature of fundamental differences among the observed explosions. 

The light curve tracks the temporal
evolution of the energy release. Most of the energy generated in the
explosion by burning matter to nuclear statistical equilibrium goes
into unbinding the white dwarf. The rest is stored in
radioactive material synthesized in the explosion (Clayton 1964, 
Colgate \& McKee
1969). The conversion of the $\gamma$-rays from the radioactive decays to
lower energy photons and their escape from the ejecta defines the light 
curve (Arnett 1982, Leibundgut \&
Pinto 1992). Since the release is highly time dependent
due to the decreasing column density and rapidly changing opacities
in the envelope, it has been very
difficult to calculate light curves of SNe~Ia. A technical
complication is the conversion of the total released energy into the
filter passbands of observed light curves requiring the calculation of 
the complete spectrum to obtain enough
wavelength points. The
observers have had little input into the bolometric light curves
of SNe~Ia (Leibundgut \& Pinto 1992). Attempts have been made by 
Suntzeff (1996) for SN~1992A
with superb and Leibundgut (1996) for SN~1990N with rudimentary
data. Only recently have data sets, like the one for
SN~1994D, become available for a detailed determination of a
bolometric flux evolution (Vacca \& Leibundgut 1996). Interestingly,
the variations among SNe~Ia persist in the bolometric light curves and
a variety of shapes is found. Most strikingly, the second maximum 
observed in the infrared appears as an inflection of variable strength 
in the bolometric light curves (Contardo et al. 1998). 
This points to important fundamental differences in the
energy release among individual SNe~Ia. The light curve correction
procedures will have to be tested on the bolometric light curves which
are a much more physical quantity than the filter light curves. 
It should be pointed out that the combination of these effects indicates
a more complicated picture than what has been captured in simple
scaling relations. They may be adequate to render SNe~Ia 
suitable objects to measure cosmological distances, but do not suffice to
describe the physics of the explosion.

Apart from the general difficulty of demonstrating the small
luminosity scatter of SNe~Ia,
their use also requires a careful error analysis. Section 2 will
outline how cosmology is done with standard candles and basically
summarizes work presented in fundamental papers of several decades
ago. In \S3 we apply the theory to observed SNe~Ia. A discussion of
the sources of systematic errors and how they are treated is
given. The analysis of the first set of 4 SN~Ia serves as an
illustrative example. We conclude (\S4) with an assessment of how suitable
SNe~Ia are for cosmology.

\section{The use of standard candles in cosmology}

The principle of standard candles is probably the simplest and most often 
used method to
measure cosmological parameters. The combination of the distance modulus
and the Hubble law at small redshifts provides a direct way
to measure the Hubble constant, $H_0$. 
The dimming of a standard candle as a function
of redshift $z$ is described by
\[m = 5\log z + 5\log\frac{c}{H_0} + M + 25. \]

Given the fixed absolute magnitude $M$ of a known standard candle 
any measurement of the apparent magnitude $m$ of an object at redshift
$z$ provides the value of Hubble's constant (in units of
km~s$^{-1}$~Mpc$^{-1}$). This is typically shown in
a Hubble diagram, $\log(cz)\, vs.\, m$.

For cosmologically significant distances, where the effects of the matter
and energy contents of the Universe become significant, the luminosity distance 
is defined by the integration over the line element along the line of sight.

All early papers on this subject used the series expansion 
\[m = 5\log z + 1.086(1-q_0)z +5\log\frac{c}{H_0} + M + 25 \]
(Heckmann 1942, Robertson 1955, 
Hoyle \& Sandage 1956, Sandage 1961). 
$q_0$ here is the deceleration of the expansion.
The integral of the line element 
can be solved analytically only in 
some specific cases (e.g. negligible cosmological constant: Mattig 1958; 
special cases including a cosmological constant: Mattig 1968). The earliest
publications (McVittie 1938, Heckmann 1942) already warned of the 
dangers involved 
in the expansion of the exact equation, and Mattig (1958) showed that for 
models without a cosmological constant a second order term makes significant 
contributions.

A modern derivation of the
relations for an expanding universe with a cosmological constant
is given in Carroll et al. (1992). 
Using the Robertson-Walker metric the luminosity distance in an
expanding universe, allowing for a cosmological constant $\Lambda$, is 
%\[D_L = \frac{(1+z)c}{H_0\sqrt{|\kappa|}} S\{\sqrt{|\kappa|} \int_{0}^{z}
%\frac{1}{\sqrt{\kappa(1+z^{\prime})^2 + \Omega_M(1+z^{\prime})^3 +
%\Omega_{\Lambda}}}\} dz^{\prime}. \]
\[D_L = \frac{(1+z)c}{H_0|\kappa|^{1/2}} \;\; S\!\left\{ \begin{array}{c} \rm 
\\ \rm \end{array} \right.\!\!\!\!\!\!\!|\kappa|^{1/2} \int_{0}^{z}
[\kappa(1+z^{\prime})^2 + \Omega_M(1+z^{\prime})^3 +
\Omega_{\Lambda}]^{-1/2} dz^{\prime}\left\} \begin{array}{c} \rm \\ \rm
\end{array} \right.\!\!\!\!\!\!\!. \]

Here $\Omega_M =
\frac{8\pi G}{3H_0^2}\rho_M$ stands for the matter content, which depends 
only on the mean matter density of the universe
$\rho_M$, and $\Omega_{\Lambda} =
\frac{\Lambda}{3H_0^2}$ describes the
contribution of a cosmological constant to the expansion factor. 
$\kappa$ is the curvature term and obeys 
\[\kappa = 1 - \Omega_M - \Omega_{\Lambda}. \]

$S(\chi)$ takes the form

\[S(\chi) = \left\{ \begin{array}{lll}
		   \sin(\chi) & & \kappa < 0 \\
		   \chi & {\rm for} & \kappa = 0 \\
		   \sinh(\chi) & & \kappa > 0.
		   \end{array}
          \right. \]

The cosmic deceleration in these models is defined as $q_0 =
\frac{\Omega_M}{2} - \Omega_{\Lambda}$. 

\begin{figure}[t]
\plotfiddle{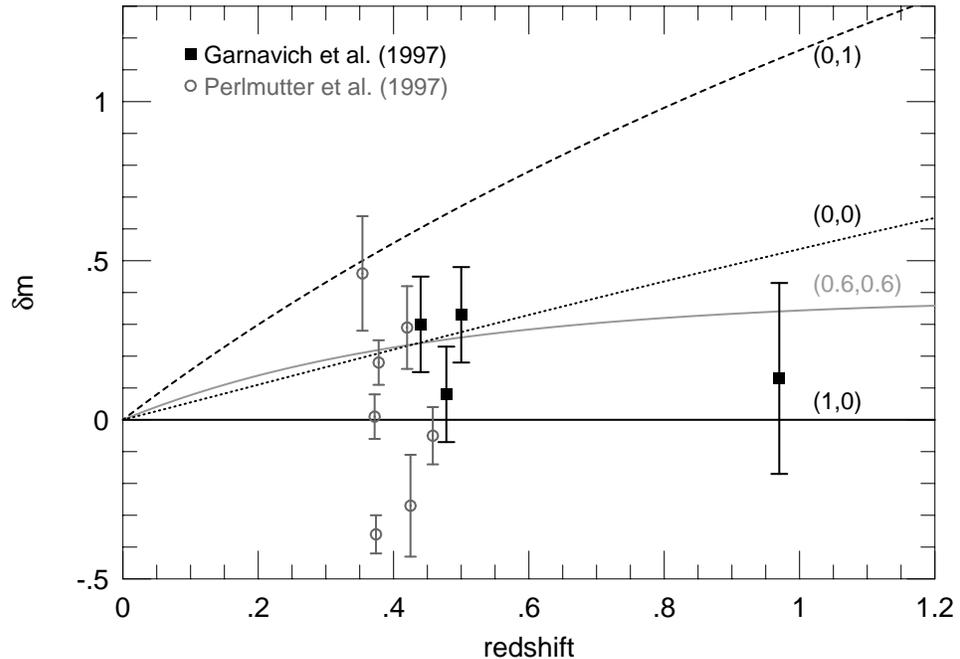}{8cm}{0}{50}{50}{-200}{-30}
\caption{Magnitude differences of a standard candle for various cosmological 
models. The lines for various combinations of ($\Omega_M,\Omega_{\Lambda}$)
are shown. They correspond to $q_0 = 0.5$ (solid line), $q_0 = 0$
(dotted), $q_0 = -1$ (dashed), and $q_0 = -0.3$ (grey).}
\label{fig:rel}
\end{figure}

The dimming of standard candles in different cosmological models is normally 
displayed as a set of lines in the Hubble diagram (Sandage 1961,
Leibundgut \& Spyromilio 1997, Perlmutter et al. 1997).
It is, however, more instructive to plot a diagram of the
magnitude differences between the various world models (Fig.~\ref{fig:rel}, cf.
also Schmidt et al. 1998, Garnavich et al. 1997). The differences between 
the various models become more apparent in this diagram. 
A standard candle in an open universe ($\Omega_M=0$, $\Lambda=0$) would appear
0.17 magnitudes fainter at a redshift of 0.3 than in a flat
universe ($\Omega_M=0.5$, $\Lambda=0$). This difference 
increases to 0.33 mag at $z=0.6$ and 0.54 mag for $z=1.0$
(Fig.~\ref{fig:rel}). These are small
values considering how difficult the observations are and the
corrections which are needed to obtain a significant measurement.

The present-day cosmic deceleration $q_0$ combines all energy sources
contributing to the change of the expansion rate of the universe. It
thus represents a fundamental parameter for the description of the
Universe we live in. For models without the cosmological constant the
fate of the Universe is encapsulated in $q_0$. There is, however, a
degeneracy in $q_0$ when a cosmological constant is included (cf.
Fig.~\ref{fig:rel}). 
The separation of
$\Omega_M$ and $\Omega_{\Lambda}$ requires the observation of standard
candles over a large range of redshifts (Goobar \& Perlmutter 1995). 

It is important to realize that
the value of the Hubble constant is not required for the
determination of $q_0$. What is important is the apparent magnitude 
difference of a standard candle measured at two different redshifts. 
Distant supernovae have to be compared to a set of nearby supernovae where 
$q_0$, or equivalently the curvature $\kappa$, 
is negligible.

\section{The cosmological deceleration as measured by SNe~Ia}

\subsection{Systematics}

Even though the scatter in peak luminosity is small, SNe~Ia 
are not perfect standard
candles. Various effects can play an important role when they are applied to
measure distances. 
Most importantly, the samples which are compared for the derivation
of the cosmological parameters have to be compatible and commensurate.

Even a perfect standard candle suffers from extinction along the
line between the object and the observer. The differences in the
absorption along different lines of sight, least known in the 
host galaxy, can modify the observed luminosities.
Supernovae must be checked carefully for reddening. 
SNe~Ia display a
range of intrinsic colors and can not be corrected for absorption based
on color alone. Rest-frame color still is a good indicator of
unreddened objects as there is a well-defined blue cutoff in the color
distribution, but it can only be measured through light curves in
at least two filters. The concentration on SNe in elliptical galaxies
could partially remedy this problem (Branch \& Tammann 1992), but a suitable
sample of nearby SNe~Ia in elliptical galaxies would have to be established. 
There also appears to be a
luminosity difference among SNe~Ia between star-forming and elliptical
galaxies which may be a worry (Hamuy et al. 1996, Schmidt et al. 1998).

Some correction methods treat reddening implicitly (Riess
et al. 1996) with the assumption that the color correlates with the
peak luminosity. Light curves in at least two filters are mandatory for this
application.

Type Ia supernovae have to be recognized as such which is
reliably done through spectroscopy (Oke \& Searle 1974, Filippenko 1997).
Peculiar supernovae have been distinguished by their spectral
appearance (Filippenko et al. 1992a, 1992b, Phillips et al. 1992, 
Leibundgut et al. 1993, Branch et al. 1993). 
Spectroscopic observations of supernovae are an indispensable requisite for
any supernova research.
Contamination of the sample by supernovae of other
types is unacceptable and dangerous. Peculiar SNe~Ia have
to be detected as well. The High-z Supernova Search (Schmidt 1997) thus has set 
out from the beginning to obtain spectroscopy of all supernovae discovered 
in its search.

Any global differences in the composition and global
properties of the comparison samples introduce systematic errors 
in the result. To
illustrate this consider the average luminosity of two samples which
are used to derive cosmological parameters. If, due to Malmquist bias,
the more distant sample has a systematically higher luminosity than
the nearby sample, and we would just apply the standard candle
paradigm, the mean distance difference would be underestimated. The
implications are the derivation of too large a value for the Hubble 
constant or an overestimation of $q_0$. 

There are two more uncertainties when using SNe~Ia for the
determination of $q_0$. One is evolution and the other distortions due
to gravitational lensing effects. Evolution would be a clear violation
of the standard candle paradigm. Since we are looking at explosions,
which occurred at a time when the universe was only half its current
age, we have to make sure that they are the same as the
ones we observe nearby. This would be a bold assumption
if unchecked. Depending on the progenitor stars of SNe~Ia there could
be substantial evolutionary effects. If SNe~Ia result from
sub-Chandrasekhar-mass white dwarfs, the mass dependence could introduce
luminosity differences which would skew the luminosity function 
(Ruiz-Lapuente et al. 1995, Canal et al. 1996). Since
more massive white dwarfs are formed first, it is conceivable that
these distant SNe~Ia are from an intrinsically more massive
population and thus also more luminous, which would have an effect
comparable to the Malmquist bias described above. 
Even with Chandrasekhar-mass
progenitors the luminosity could be substantially changed due to
abundance differences in the parent populations. The smaller
metallicities could mean that the line opacities governing the
radiation escape in SNe~Ia are significantly changed which would
result in changed luminosities in the observed filters. The only way
to check any of the above conjectures lies in carefully observing the distant
objects and asserting their similarity with the nearby sample. If the
distant sample looks the same as the nearby one in all aspects we 
can measure, we have a
good chance that they have the same average luminosity as well. It is not
sufficient to exclude strange-looking or non-conforming objects as they may 
be the sign of
real differences invalidating the standard candle approach
altogether.

Gravitational lensing always affects light traveling over cosmological
distances. Amplifications introduced by the
lensing have been investigated in combination with
SNe~Ia (Wambsganss et al. 1997). Gravitational lensing, unlike absorption, is
achromatic. It can not be inferred from the observed light alone.
Luckily, at the redshifts of the supernova
searches ($z < 1$ so far) the systematic amplification (or dimming) of
standard candles for current models of the mass distribution is small 
($\Delta q_0 < 0.04$ for all $z<1$; Wambsganss et al. 1997).

\subsection{Observing distant supernovae}

\begin{figure}
\plotfiddle{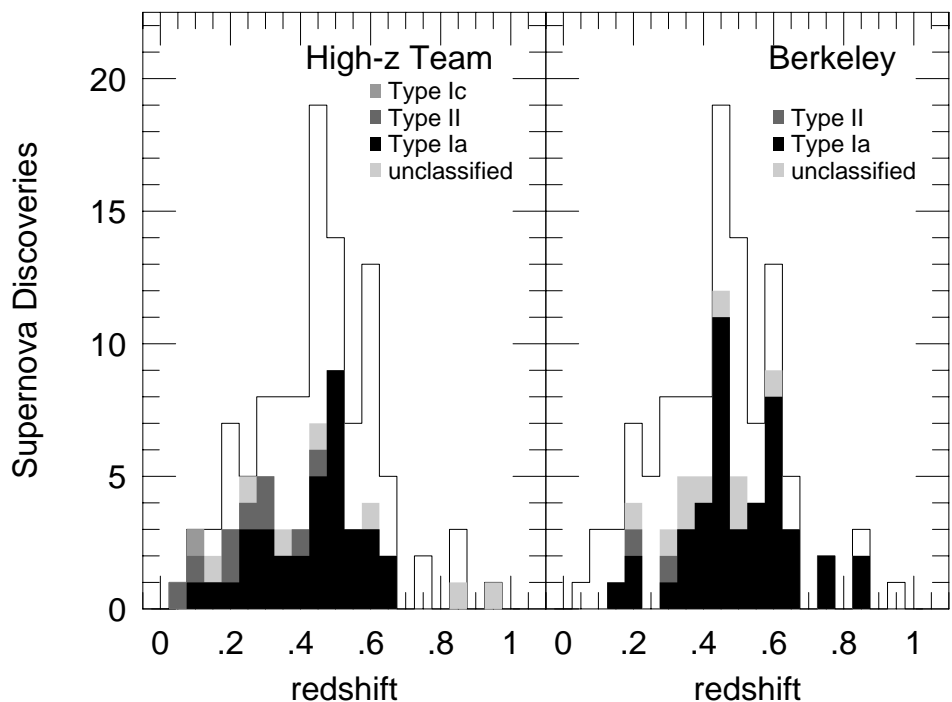}{8cm}{0}{75}{75}{-230}{-252}
\caption{Supernova discoveries reported by the distant SN searches
in IAU Circular until July 1997.
The top line indicates the total number of SNe from both searches
combined.}
\label{fig:snehist}
\end{figure}

There are currently two large observational programs which try to 
determine the deceleration of the universe using SNe~Ia. 
They both make use of SNe~Ia as
standard candles or modified versions thereof. The 
High-z Supernova Search Team
(Schmidt 1997, Schmidt et al. 1998, Garnavich et al. 1997,
Leibundgut et al. 1996, Leibundgut \& Spyromilio 1997) and the 
Berkeley Cosmology Project (Perlmutter et al. 1995, 1997)
have discovered a large number of supernovae over the last three
years. In fact, these two supernova searches now contribute the majority
of all discoveries per year.
Figure~\ref{fig:snehist} shows a summary of all SNe~Ia found in
these two searches up to June 1997. 
It should be compared to a similar figure from
1996 (Leibundgut \& Spyromilio 1997) to appreciate the rapid progress. 
Both groups have mastered the art of
detecting these extremely faint objects very efficiently. 
The numbers reported are actually lower
limits as both groups now only announce bona fide supernovae at high
redshifts. 
Not even all types of SNe are reported any longer as the highest
interest is in distant SNe~Ia.
This should be
taken as a word of caution when interpreting the numbers for
supernova rates. The discovery of SNe~Ia dominates the distribution.
Only very few Type II supernovae have been announced at redshifts larger 
than 0.3
and a single SN~Ic has been discovered in the High-z SN Search.

The total number of SNe discovered in the two searches is 107 objects, 70 of
which are confirmed SNe~Ia at $z>0.25$ (29 High-z SN Search and 41 Berkeley
Cosmology Project).

The emphasis has shifted from
actually finding the supernovae to organizing the critical follow up
observations. 
Spectroscopy is the prime classification resource and also serves as a
discriminator against
unusual objects. Photometry is required to establish the essential
light curve parameters, if such a correction should be applied. 

These observations are challenging for any telescope.
As an example consider a SN~Ia at z=0.5: the observed peak
magnitude reaches 21.9 in the $R$ band (SN~1995K; Schmidt et al. 1998), but 
the crucial light curve parts for the decline rate determination are
around $R=23$ and, if some of the light curve tail should be observed,
magnitude levels with $R>25$ have to be considered. It can not be
emphasized enough how important it is to achieve accurate photometry
at these magnitudes as only small deviations can result in substantial, in
terms of the effect to be measured, over-corrections. This is even
more so, should systematic effects creep in, which skew the correction
unilaterally through, e.g., contamination by the galaxy background.
It is dangerous to ``blindly'' apply
light curve correction procedures without the assurance that the objects are
indeed SNe~Ia.

All observations obtained so far indicate that the distant supernovae
look very similar to the local ones (Leibundgut et al. 1996,
Leibundgut \& Spyromilio 1997, Perlmutter et al. 1997, Schmidt et al.
1998, Garnavich et al. 1997).   Spectra, light curves, and rest-frame
colors of distant supernovae appear identical to the bulk of the
local population, and evolutionary effects are not apparent (Schmidt et
al. 1998). 

Technical issues in the analysis are the accuracy of the photometry,
the calculation of K-corrections, any application of luminosity
corrections, and uncertainties in the zero-point defined
by the nearby sample. While the photometry and the K-corrections appear
to be under control, the correction for the light curve shape can
become problematic. The maximum magnitude of a SN~Ia is measured
fairly easily as the techniques employed in these searches
by design discover SNe close to maximum
(Perlmutter et al. 1997, Schmidt et al. 1998). Typical photometric
errors of individual light curve points are about 0.05 to 0.15
magnitudes near the peak, but increase to 0.3 magnitudes a couple of weeks from
maximum. The error in the determination of the filter maximum is around 0.1
to 0.2 magnitudes. The uncertainties in the K-corrections are
negligible ($<$0.05 magnitudes; Kim et al. 1996, Schmidt et al. 1998). 
Light curve corrections introduce two uncertainties. 
They are defined by photometry points which are less well
measured than near maximum and the correction parameters themselves carry
uncertainties (Hamuy et al. 1996). It has yet to be shown
unambiguously that the light curve corrections also decrease 
the scatter in the distant sample. Contrary to the simple
determination of the maximum magnitude these methods make use of the
whole light curve shape. This is a very valuable asset when the
photometry errors are not significant, but can become problematic when
applied to points with large errors. In addition, the corrections
change the zero-point of the sample, a
very contentious issue in the Hubble constant discussion (Hamuy et
al. 1996, Saha et al. 1997). It will be very important to resolve this
ambiguity also for the determination of $q_0$. 

The statistical uncertainty in the
zero-point of the nearby SN~Ia sample is typically about 0.03 
(Hamuy et al. 1996). The
contribution to the uncertainty in $q_0$ scales with $1/z$ and for a
sample with a mean redshift of 0.5 contributes about 0.06 to the total
uncertainty in $q_0$.

\subsection{The first four High-z supernovae}

SN~1995K is the first SN~Ia discovered in the High-z SN Search (Leibundgut
et al. 1996, Schmidt
et al. 1998). Three more supernovae have become available recently
(Garnavich et al. 1997). We will use this small sample of four SNe to 
exemplify the analysis of distant SNe~Ia and describe the
systematic influence of the basic assumptions.

The spectrum and the light curves of SN~1995K show no signs of
any peculiarity for a SN~Ia at $z=0.478$
(Leibundgut et al. 1996, Leibundgut \& Spyromilio
1997, Schmidt et al. 1998). Once
the time dilation has been removed the light curve is comparable to
regular nearby SNe~Ia. K-corrections are nearly
constant throughout the observed phase due to the near match of the $R$
and $I$ filters to rest-frame $B$ and $V$, respectively, at $z \approx 0.5$.
The rest-frame color derived for
this object is $(B-V)\approx -0.1$, consistent with unreddened 
nearby objects (Vaughan et al. 1995, Tammann \& Sandage 1995). 
The peak magnitude in rest-frame $B$ and $V$ is 22.9 and 23.0,
respectively. It can thus be assumed that it is a good candidate 
for the measurement of deceleration. 

SN~1997ce and SN~1997cj are spectroscopically confirmed SNe~Ia at
redshifts of 0.44 and 0.50, respectively (Garnavich et al. 1997). 
Spectra of these two supernovae show the characteristic 
Si~II (SN~1997ce) and Ca~II (SN~1997cj) absorption.
They both have a well sampled light curve from ground-based and $HST$
observations in rest-frame $B$ and $V$ without apparent peculiarities. The
color at maximum for both objects is also consistent with $(B-V) \approx
0.0$ indicating that absorption is
negligible. We find
peak magnitudes of 22.9 ($B$) and 23.0 ($V$) for SN~1997ce and 23.3 ($B$ and
$V$) for SN~1997cj from simultaneous fits to light curve templates.

The most distant supernova in the sample, SN~1997ck, is at a redshift of
$z = 0.97$ as deduced from a spectrum which shows a single, narrow
emission line, identified as [O~II], from the galaxy. The time dilation in 
the light curve indicates a redshift of $1.1 \pm 0.2$ for the best fitting
light curve of a nearby SN~Ia (SN~1991T).
At this redshift the $V$ band is shifted beyond 1$\mu m$
and no data could be obtained. Hence we have no indication whether dust
absorption is affecting this object. The following analysis of the $B$ 
light curve assumes negligible absorption. 

Figure~\ref{fig:rel} indicates the position of these supernovae in comparison
with nearby SNe~Ia (Hamuy et al. 1996). 
$B$ and $V$ observations have been combined for the points of the High-z
sample.
This diagram has the advantage that both filter measurements
can be combined in the same graph. This also means that the observations
in two filters, apart from their necessity for accurate photometry and
the check for absorption, increases the number of observations in the
data set thus constraining the uncertainty. 
The figure also displays 
the supernovae reported by the Berkeley group (Perlmutter et al. 1997). 

The diagram in Figure~\ref{fig:rel} assumes a perfect standard
candle, i.e. no light curve shape correction has been
applied. It uses the zero-point given in equations
4 and 5 of Hamuy et al. (1996). There is a large overlap between the two
data sets, but a clear lack of bright supernovae in the High-z sample.
We find values for $\Omega_M$ and $\Omega_{\Lambda}$ by $\chi^2$ fitting.
Only for the uncorrected $B$ data do we find an acceptable solution
($\chi^2$ minimum), if the
cosmological constant is ignored. The best value in this case is
$\Omega_M = 0.1\pm0.6$. The best solution is found for $\Omega_M
= 0.6$ and $\Omega_{\Lambda} = 0.6$ (corresponding to $q_0 = -0.3$)
with a very large uncertainty in
both parameters. For the $V$ filter data we find no solution with
$\Omega_{\Lambda} = 0$ and $\Omega_M \geq 0$. The best solution lies
near $\Omega_M=0.5$ and $\Omega_{\Lambda}=0.7$ ($q_0 = -0.45$). 
The original result for the Berkeley set was close to $\Omega_M = 0.9$,
$\Omega_{\Lambda}=0.1$ (Perlmutter et al. 1997).

\begin{figure}[t]
\plotfiddle{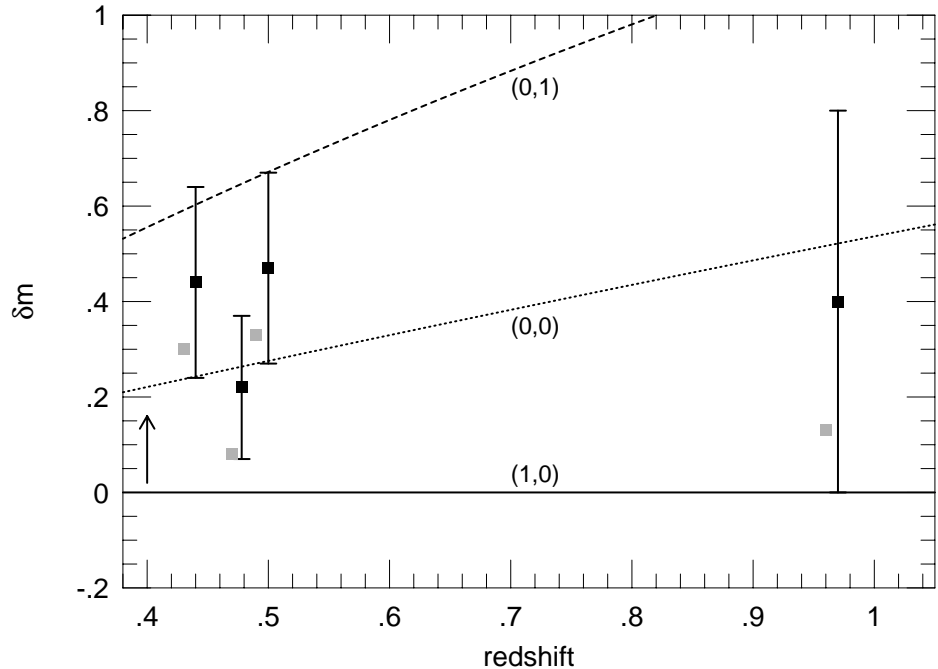}{8cm}{0}{50}{50}{-200}{-30}
\caption{Same as Fig.~\ref{fig:rel} but after application of the light
curve correction. The arrow indicates the shift due to the zero-point
change (0.14 magnitudes). The faint dots indicate the former position of
the data( shifted in redshift for clarity). The errors have increased slightly
due to the additional uncertainty of the light curve fitting.
}
\label{fig:sys}
\end{figure}

The picture changes, when we apply the correction for the
light curve shape. The model lines shift by 0.14 magnitudes due to
the change of the zero-point provided by the local sample 
(Hamuy et al. 1996; equations 7 and
8). At the same time we have to
correct the distant supernovae for their light curves. With the
exception of SN~1997ck, which seems to have been similar to SN~1991T in
light curve shape, all SNe are best approximated by the template
light curve of nearby SNe~Ia (cf. Garnavich et al. 1997). Thus, any light
curve corrections for these objects are small ($<0.1$ mag). The
correction of the $B$ maximum brightness of SN~1997ck is 0.13
magnitudes assuming it was identical to SN~1991T.

The global effect is illustrated
in Figure~\ref{fig:sys}. The arrow indicates the shift of the data points due
to the change of the zero-point of the lines. 
After the correction all points scatter around the line for an open
universe ($\Omega_M=0, \Omega_{\Lambda}=0$). A $\Lambda$-dominated 
universe is now favored with best fits of 
$(\Omega_M, \Omega_{\Lambda}) = (0.3,0.6)$ for the $B$ data and
$(\Omega_M, \Omega_{\Lambda}) = (0.3,0.9)$ for $V$. These values
correspond to $q_0 = -0.45$ and $q_0 = -0.75$, respectively.
The value for $\Omega_M$ has decreased in the $B$ solution, because 
SN~1997ck, being identified as overluminous, has undergone a larger 
correction than the other supernovae. The photometric uncertainty in 
the measurement of this high-z supernova, however, reduces its contribution in
the $\chi^2$-fit. Thus the signal is produced mainly by the 3
supernovae near $z \approx 0.45$ and not by SN~1997ck, as is obvious
from the fits to the $V$ data. Note that any
reddening of SN~1997ck would make this supernova even brighter.
We have to emphasize that 
the number of objects is still small and it is certainly premature to 
draw definite conclusions. 

Since we do not have access to the individual photometry
of the Berkeley supernovae we can not perform the same analysis on
their data. It should be pointed out, however, that two of their SNe
lie outside the observed range of decline rates and have been excluded in their
analysis. The correction of the remaining objects nearly cancels the offset
of the zero-point shift as they arrive at almost the same result for the 
uncorrected and the corrected sample. This is surprising considering the 
effect of the zero-point alone (Fig.~\ref{fig:sys}).
A direct comparison of the two data samples is not possible, but we can 
compare SNe which have similar redshifts. 
SN~1992bi had almost the same redshift ($z=0.458$) as SN~1995K
and undergoes a large correction, but was excluded from the subsequent
analysis, because it displayed a slower decline rate than observed for 
any local SN~Ia (Perlmutter et al. 1997). 
More interesting might be the comparison of SN~1997ce with SN~1994al 
($z=0.420$) and SN~1994G ($z=0.425$). SN~1994G has a spectrum identifying the 
object as a genuine SN~Ia, while the galaxy spectrum of SN~1994al is 
described as consistent with an elliptical or S0 galaxy, which would identify
this object also as a SN~Ia. The observed rest-frame $B$ magnitudes at peak are
22.8 (SN~1994al) and 22.3 (SN~1994G). These have to be compared to the peak
magnitude of SN~1997ce which is 22.9. Thus, we have a very good agreement 
between SN~1994al and SN~1997ce, but SN~1994G deviating by about 0.5 
magnitudes. Even after corrections for the light curve shape the close
agreement between SN~1994al and SN~1997ce remains, and SN~1994G still 
is an outlier (Table~1 of Perlmutter et al. 1997). 
It is puzzling to find that our result changes so
dramatically (almost entirely due to the shift of the zero-point of the local
sample), while the analysis of Perlmutter et al. (1997) did not
result in a similar signal when they applied their light curve correction 
method.

\section{Conclusions}

The regular behavior of SNe~Ia has already provided proof of another
prediction of models of an expanding universe. 
Time dilation contributes significantly to the
filter light curves and the spectral evolution of distant supernovae
(Wilson 1939, Tammann 1979). The light curves of SN~1995K are
dilated exactly by the factor $(1+z)$ predicted for an expanding
universe (Leibundgut et al. 1996). 
The Berkeley data set has been investigated for time dilation with similar
conclusions (Goldhaber et al. 1997). 
In cases like SN~1997ck, where there is the detection of 
a single line in the galaxy spectrum, the time dilation can be
used for consistency checks, although one has to assume that the object 
indeed is a SN~Ia. 
Retarded spectral evolution, consistent with time dilation, has been 
observed in SN~1996bj at a
redshift of 0.574 (Riess et al. 1997). Nevertheless, there remain proposals
of static universe models which conform with the observed time dilation
(Narlikar \& Arp 1997, Segal 1997). 

With sufficient control the difficulties of SNe~Ia to measure the current
state of the deceleration can be overcome. Light curves in two 
filters and
spectroscopy near maximum light can identify extinction or peculiar SNe~Ia,
which may prevent the correct application of the standard
candle paradigm. Only a very careful characterization of all measurable 
aspects of
the sample of distant supernovae and its comparison with the nearby
set can lead to a successful measurement of $q_0$. The influence of
uncertainties introduced by the local sample should not be
underestimated. They are sizeable contributions to the error with
which $q_0$ can be determined. 

The largest systematic uncertainty is the application of
luminosity corrections. This is a rather intricate problem which is
discussed in connection with SNe~Ia based determinations of $H_0$, 
but applies equally to
$q_0$. Our analysis of the small SN sample available shows that the
light curve corrections produce a different result than
a simple comparison of peak magnitudes. For SNe~Ia with light curve
shapes deviating only slightly from the average this effect is entirely
in the offset of the zero-point of the local supernova sample. 
For SNe~Ia which are overluminous 
as indicated by their light curves, the standard 
candle paradigm only produces an upper limit for $q_0$. 
The existing data at this point clearly favor an open universe and even 
a contribution of a cosmological constant, although the errors are still
considerable. 

As discussed the difference between an empty
and a flat universe is 0.54 magnitudes at $z=1$
(Fig.~\ref{fig:rel}). This means
that the influence of the zero-point decreases and
a measurement at larger redshifts is less sensitive to these
systematics. The difficulty will be to obtain the accurate
photometry to reliably measure the light curve shape and apply the
corrections. The High-z SN Search has already discovered and followed 
supernovae at these redshifts. We confirm in our
analysis the result presented in Garnavich et al. (1997).
The distant objects are essential to measure
$\Omega_M$ and $\Omega_{\Lambda}$ independently (Goobar \& Perlmutter 1995).

Due to Malmquist bias one is more likely to overestimate
$q_0$ than to find too small a value. On the other hand, any residual
photometry errors which would tend to widen the light curves would lead
to the adoption of too bright a luminosity for the supernovae. 

Type Ia Supernovae may be the closest we know to a cosmological 
standard candle, yet the magnitude differences for different dynamical
states of the universe are very small
and a tight control of the samples is indispensable. In
principle all the required information can be gathered by the observations.
The knowledge of the progenitor systems and the exact radiation
physics are pivotal ingredients still missing from our picture
of these explosions (Branch et al. 1995). 
With such an understanding the inference of the
standard candle character and possible evolutionary scenarios can be
tested. Until then we have to rely on the control provided by 
careful observations and analysis.

\acknowledgments

I am grateful to Gustav Tammann for directing my early research into
supernovae and the never failing support he has given me since I first 
walked into his office. The data discussed in this paper have been
collected by the High-z SN team. I am thankful to the team members,
especially Brian Schmidt and Peter Garnavich, on the work of the
photometry reported in this paper.

\end{document}